\documentclass[12pt, journal, letterpaper]{IEEEtran}
\usepackage{amsmath}
\usepackage{graphicx,subfigure,comment}
\usepackage[table]{xcolor}
\usepackage{url} %for \path
\usepackage{dsfont}
\usepackage{cite}
\usepackage{tabularx}
\usepackage{dblfloatfix}
\usepackage{enumitem}

\newcommand{\Fig}[1]{Fig.~\ref{fig:#1}}

\newcommand{\Sec}[1]{Sec.~\ref{sec:#1}}

\setlist[itemize]{leftmargin=4mm}
\setlist[enumerate]{leftmargin=5mm}

\begin{document}

\title{Federated Learning at the Network Edge:\\When Not All Nodes are
Created Equal} \author{Francesco Malandrino, Carla Fabiana Chiasserini}
\maketitle

\begin{abstract} Under the federated learning paradigm, a set of nodes
can cooperatively train a machine learning model with the help of a
centralized server. Such a server is also tasked with assigning a weight
to the information received from each node, and often also to drop
too-slow nodes from the learning process. Both decisions have major
impact on the resulting learning performance, and can interfere with
each other in counter-intuitive ways. In this paper, we focus on edge
networking scenarios and investigate existing and novel
approaches to such {\em model-weighting} and {\em node-dropping}
decisions. Leveraging a set of real-world experiments, we find that
popular, straightforward decision-making approaches may yield poor
performance, and that considering the quality of data in addition to its
quantity can substantially improve learning. 
\end{abstract}

\section{Introduction} \label{sec:intro}

Federated learning (FL) is a distributed machine learning paradigm
whereby a set of {\em learning nodes} cooperate in training a model
(e.g., a neural network) with the assistance of a centralized {\em model
server} and without the need to share their local data. 
FL has been introduced~\cite{konen2015federatedOptimization} in 2015 by
Google, with the goal of leveraging the computational power of end-user
devices -- most notably, smartphones -- without the privacy and security
concerns arising from sharing the potentially sensitive information they
own. As discussed in \Sec{fl}, it has since been widely adopted in
edge computing scenarios, owing to its ability to
blend device- and server-based computation, and to enable cooperation
between devices regardless of their location.

\begin{figure}[t] \centering
\includegraphics[width=.85\columnwidth]{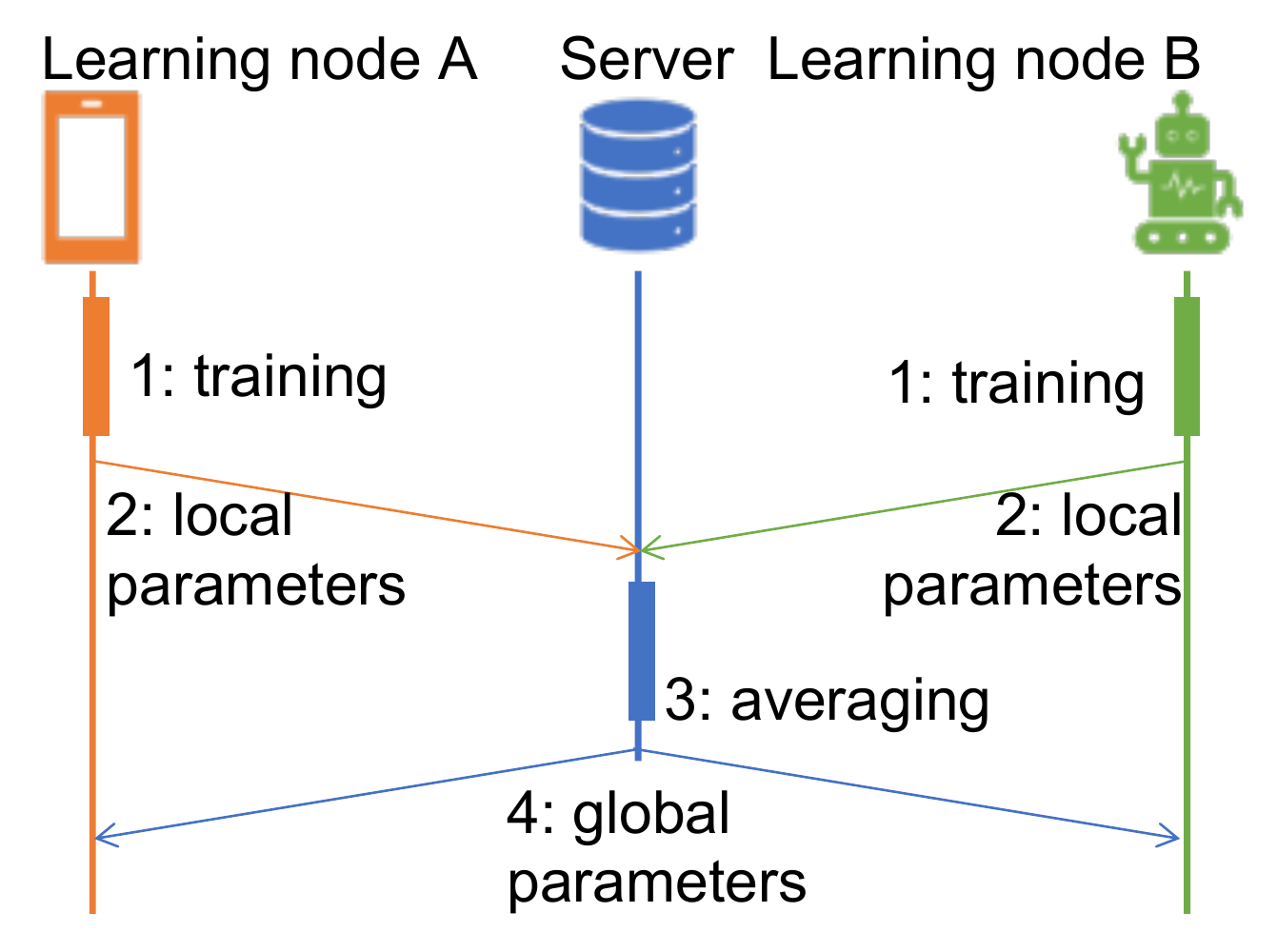} 
\caption{Main steps of each 
iteration of the federated learning paradigm:
learning nodes train their local model (1) and send the local
parameters to the server (2); the server performs a weighted averaging
of the model (3) and sends the global parameters back to the learning
nodes (4). 
\label{fig:sequence} } %caption 
\end{figure}

FL includes the following high-level steps, summarized in
\Fig{sequence}: 
\begin{enumerate} 
\item learning nodes train a local
model based on local, on-device data; 
\item learning nodes send the model
parameters -- and nothing else -- to the server; 
\item the server combines the parameters coming from different learning nodes; 
\item the server sends the combined, {\em global} parameters back to the learning nodes; 
\item the learning nodes replace their local parameters with the global ones, and move to
a new training iteration (step 1),
till the desired accuracy is achieved.
\end{enumerate} 
Step~3 usually takes the
form of a {\em weighted averaging} of the local 
parameters~\cite{konen2015federatedOptimization,jointframework}. 
Weights are
assigned to individual learners and reflect the magnitude of their
(expected) contribution to the learning process; indeed, as discussed
next, properly assigned model weights is one of the main decisions
learning servers can make.

Originally envisioned for homogeneous nodes dealing with homogeneous datasets
(e.g., in a classification problem, datasets adequately representing all
classes), FL  has also the potential to deal with {\em heterogeneity},
both in the capabilities of learning
nodes~\cite{kang2020reliable,jointframework} and in their local
data~\cite{infocom20-noniid,jointframework}. In both cases, the key is
to endow the model server with additional responsibilities: the {\em
weights} given to different local models in the averaging phase (step 3
in \Fig{sequence}) can account for the quantity of data they are trained
upon; at the same time, if some nodes are consistently slower than
others, they can be {\em dropped} from the learning
process~\cite{kang2020reliable,client-selection}.

Our key observation is that these two decisions, {\em model-weighting}
and {\em node-dropping}, are deceptively simple, and their impact on the
overall learning process is often misunderstood and underestimated.
Many state-of-the-art works take straightforward approaches to these decisions, 
which may result in poor performance under richer, more complex scenarios. 
Our contribution is therefore to shed a light on the model-weighting and node-dropping decisions, 
studying how they should account for the quantity {\em and quality, e.g., variety,} 
of data available to local nodes, as well as for the data processing time.
In so doing, we  focus on an edge network scenario and 
leverage a set of experiments using
the popular \path{tensorflow} library and the recent
Fashion-MNIST dataset.

% We find that choosing the right model-weighting strategy can also help node-dropping decisions, and that the diversity of data shall be given more importance than their quantity.

In the remainder of the paper, we describe federated learning in edge
scenarios and the associated challenges in \Sec{fl}, before narrowing
the focus on model-weighting and node-dropping strategies in
\Sec{strats}. We then describe our experiments and results in \Sec{experiments}, and
the main lessons learned in \Sec{takeaway}, along with pointers to
further promising research directions. Finally, \Sec{conclusion}
concludes the paper.

\section{Federated Learning in Edge Scenarios} \label{sec:fl}

Edge computing is a distributed paradigm
predicated upon performing the computation as close as possible to the
user nodes requesting it, i.e., at servers located at the {\em edge} of
the network infrastructure. It also includes scenarios where user nodes themselves have
computational and/or storage capabilities, and require edge support for
coordination, or to offload the heaviest computation tasks. FL has
long been identified as an excellent match for edge computing scenarios, and many
research works aim at making it in such scenarios as efficient as
possible.
At the same time, {\em communication} is a major issue for FL in edge scenarios. Nodes can be connected with the edge-based server in many ways and through different technologies; thus, their connectivity has a major impact on the latency incurred when sending model updates -- and, indeed, on whether or not such updates are received in the first place.

Specifically, as detailed in~\cite{in-edge-ai},
edge computing is more effective than fully-distributed, device-to-device
networks at tackling the main factors hindering the performance of FL, namely, the
different node capabilities, available data, and
unpredictable communication delays and shortages.
Narrowing its focus to node capabilities, \cite{client-selection}~aims
at choosing the set of learning nodes that results in the shortest
learning time, solving a double-edged conundrum. On the one hand, more
nodes mean that convergence can be reached in fewer iterations; on the
other hand, the duration of each iteration is determined by the slowest
node~\cite{neglia}. In a similar spirit, \cite{jointframework}~addresses
the problem of jointly selecting the learning nodes to use for the
learning process and assigning them the wireless resources they need to
communicate effectively. Setting in a fully-decentralized {\em fog}
scenario where no learning server may be present,
\cite{infocom20-fog}~tackles many issues relevant to edge computing, including
device mobility and the possibility of offloading computation from a
node to another.

Shifting the focus towards the major issue of communication between FL nodes and server,
the authors of
\cite{robust-efficient}~seek to reduce the communication overhead of FL
by proposing a compression algorithm suited for federated learning
settings. Their algorithm outperforms existing schemes when local
datasets are heterogeneous, thus making the high-frequency communication
required by FL viable in low-bandwidth scenarios. Heterogeneous datasets
are also identified as a major problem in~\cite{infocom20-noniid}, which
envisions extracting a homogeneous subset from each local dataset in
order to avoid bias and training errors.
\begin{comment}
In a similar spirit,
\cite{jeong2018communication}~proposes to both {\em augment} local
datasets by generating new samples via a generative model, thus making
them homogeneous, and reducing the overhead by aligning ({\em
distilling}) the models being trained, so that periodic updates can be
dispensed with.
\end{comment}

Following an orthogonal, more theoretical approach, several
works~\cite{neglia,wang2019adaptive} aim at characterizing the learning
performance, deriving closed-form expressions for their (expected)
training time. Such a characterization is then exploited to make optimal
or near-optimal decisions on the cooperation among nodes~\cite{neglia}
and the equilibrium between local learning and global
updates~\cite{wang2019adaptive}. In order to obtain manageable
closed-form expressions, some of these works make simplifying
assumptions (e.g., that local datasets be homogeneous), or target
specific parameter optimization algorithms (e.g., stochastic gradient
descent).

One scenario where nontrivial model-weighting decisions are routinely used is {\em asynchronous} 
FL~\cite{lu2019differentially}, where nodes may join the learning process at different times and  
model-weighting serves the purpose of quickly including newly-arrived nodes in the learning process. 
In our case, the purpose is different, namely, to adapt model weights to the contribution each 
node can give to the overall learning process, and weed out those nodes that may 
have a negative impact on the  learning performance.

\section{Model-weighting and Node-dropping Strategies in Federated
Learning} \label{sec:strats}

Model-weighting and node-dropping are the most fundamental decisions the
learning server can make, and arguably among the simplest to enact. At
the same time, as discussed in the following, these decisions can be
leveraged to address all the main issues of FL, either in combination
with the strategies reviewed in \Sec{fl}, or as an alternative to them.

\paragraph*{Insufficient quantity of data}

%It is a well-known feature of neural networks that they require
%significant quantities of data for training.
In many FL scenarios, some
learning nodes may not have enough
%a sufficient quantity of
local data, thus
being unable to properly train their local models;
%Besides the
%straightforward solution of collecting (or, in the case of sensors,
%waiting for) more data,
in this case,
a popular solution is {\em augmenting} local
datasets. As an example,
%\cite{jeong2018communication}~envisions
%generating additional, synthetic data samples by leveraging information
%from the other nodes, while 
the authors of~\cite{shin2020xor} propose to
combine actual data samples from other learning nodes in a
privacy-preserving way, and adding them to the local dataset.

In both cases, data augmentation is able to increase the quantity of
data available to learning nodes, without jeopardizing FL's privacy
properties. On the negative side, it may increase the complexity of the
system and its overhead; furthermore, the augmented samples come from
processing of already-existing ones, hence, do not increase the total
quantity of information.

Model-weighting decisions represent an additional, simpler way to deal
with learning nodes with small local datasets. The basic
idea~\cite{konen2015federatedOptimization,jointframework} is that model
weights shall account for the quantity of data local models are based
upon; however, as we will demonstrate next, also accounting for the
variety of such data yields even better results. In both cases, relying
on model-weighting to tackle insufficient dataset sizes has the benefit
of reducing complexity and avoid tampering with the nodes' own data.

\paragraph*{Non-homogeneous data}

When it comes to training machine-learning algorithms, the quality of
data is as important as its quantity.
There is no universally-acknowledged definition of data quality, as it is scenario- and application-dependent. For classification applications, a high-quality dataset is expected to adequately represent all existing classes, so that the classifier can be properly trained. In this sense, quality can be expressed as the number of classes existing in a given dataset or, more formally, through entropy~\cite{plos-entropy}.

On the other hand, non-i.i.d. data, where classes are under- or over-represented, is universally characterized as low-quality, and
\begin{comment}
A very important aspect of data
quality is that they represent all aspects of the phenomenon to study,
e.g., in a classification problems, that all labels are represented
proportionally to their frequency -- or, more technically, that they can
be seen as i.i.d. (independent, identically distributed) samples of the
original data. In many FL scenarios this is not the case; indeed, some
or all learning nodes may only have a partial view of the environment,
e.g., only observe some labels.

Non-i.i.d. data
\end{comment}
has immediately been identified as one of the primary
threats to successful FL. %, and targeted with several different strategies.
In addition to the augmentation approaches described
earlier~\cite{shin2020xor}, several works propose
training the local model on a subset of the local
data~\cite{infocom20-noniid}, chosen in such a way to be i.i.d.
%Counterintuitive as it may sound, in many cases using fewer data
%actually improves the learning performance.
%
%On the negative side, ignoring some data means reducing the total amount
%of information used for learning.
An alternative to ignoring data is allowing all
learning nodes to use all their data, and then weight their local models
accounting for the quality of such data. Examples of this approach
include entropy~\cite{plos-entropy}, but simpler approaches, e.g.,
counting the labels observed, can yield similarly good performance.

\paragraph*{Nodes with different capabilities}

As with other distributed learning schemes, in FL it is possible to
proceed from an iteration to the next one only when {\em all} learning
nodes have sent their local models, i.e., have performed step~2 in
\Fig{sequence}. It follows that the pace of the learning process as a
whole is determined by the {\em slowest} learning node, which becomes an
issue when different learning nodes take very different times to perform
their iterations~\cite{neglia}. Owing to the limited amount of control
that can be exerted on FL nodes, the most viable solution is often to
exclude overly-slow nodes from the learning
process~\cite{neglia,jointframework,client-selection}.

However, making such node-dropping decisions solely on the basis of
their response times may actually hurt the learning process; indeed,
longer response times can be associated with larger, higher-quality
local datasets, hence, with the nodes that may contribute the most to
the learning. A way to decrease the likelihood of this unwanted outcome
is to consider additional aspects in making node-dropping decisions,
e.g., the quantity and quality or variety of local data. By so doing, it
is possible to differentiate between nodes that are slow due to limited
capabilities (or poor
connectivity~\cite{jointframework,client-selection}) and those that have
simply more data to process.

\section{Experiment Design and Results} \label{sec:experiments}

In this section, we demonstrate how model-weighting and node-dropping
decisions can deal with heterogeneity in the quality and quantity of the
local datasets at learning nodes. Using the set of real-world
experiments described in \Sec{sub-setup}, we obtain the results described
in \Sec{sub-results}.

\subsection{Experiment setup} \label{sec:sub-setup}

{\bf Dataset and neural network structure.}
Fashion-MNIST is a dataset released by Zalando
research and aimed at providing a more challenging, drop-in replacement
for the classic MNIST handwritten digits dataset.
\begin{comment}
It includes 70,000
grayscale, 28-by-28 pixel images of articles of clothing, each
associated with one of ten labels, from ``ankle boot'' to ``pullover''.
Choosing the Fashion-MNIST dataset represents a good balance between
having a sufficiently-challenging learning task (which is not as simple
as with the original MNIST) and obtaining relevant and reproducible
results, which may be harder using less common, domain-specific
datasets.
\end{comment}
Owing to the relative simplicity of the dataset, we use a relatively
small neural network for classification. Specifically, we create a dense
network with four layers, with sizes $[28^2,200,100,200]$~neurons
(notice that the size of the first layer must match the size of the
input, i.e., $28\times 28$~pixels). Neurons use the \path{softmax}
activation function, and parameters are optimized using stochastic
gradient descent (SGD), with a learning rate of~$10^{-2}$. The network
is implemented using the popular \path{tensorflow} library, originally
developed by Google.

\begin{figure} 
\centering
\includegraphics[width=.95\columnwidth]{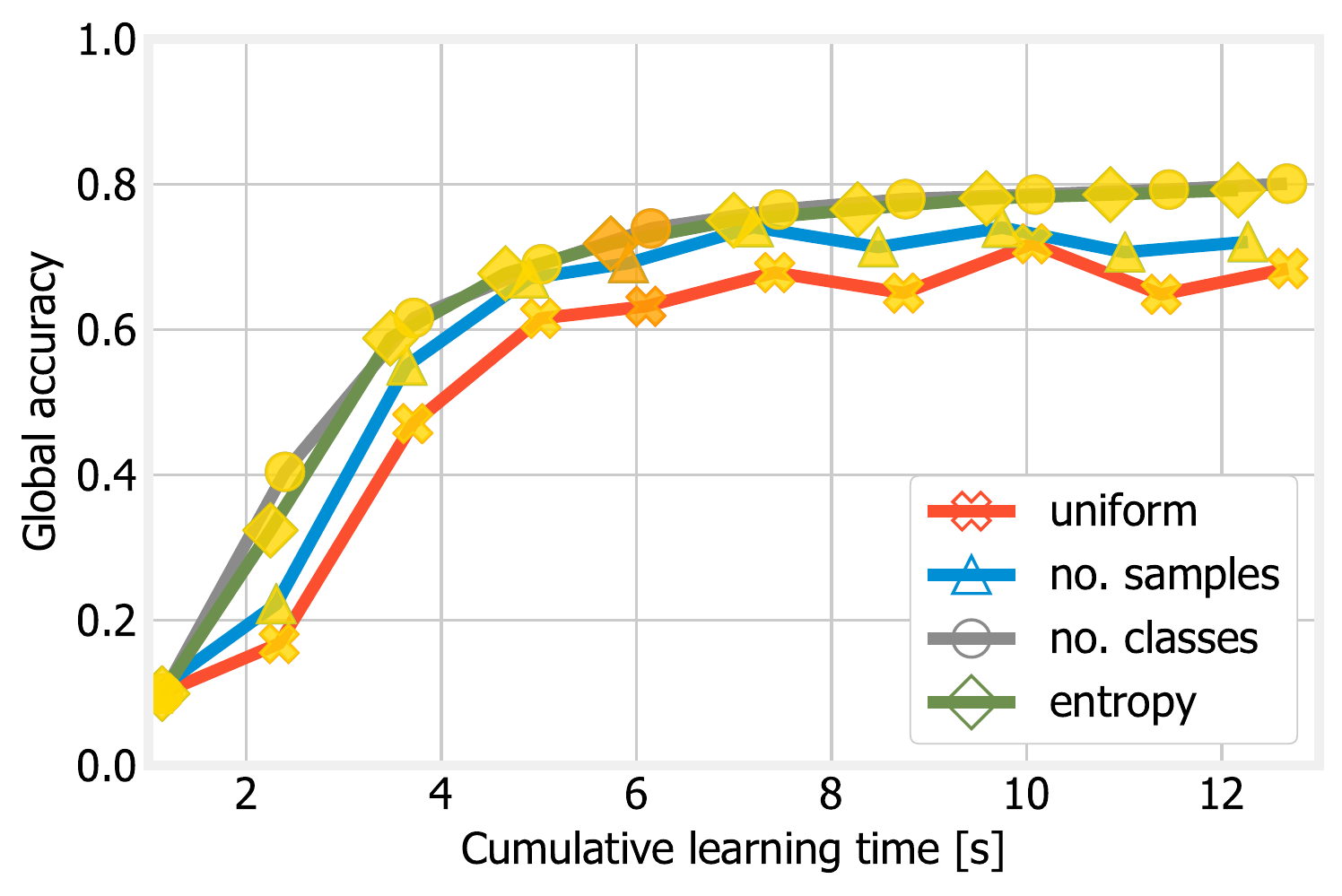}
\caption{Relationship between elapsed time and global accuracy for
different model-weighting strategies, when no nodes are dropped from the
learning process.
The color of markers corresponds to the category 
of the slowest node in that particular iteration 
(gold: yellow, bronze: orange, no other category appears).
\label{fig:learning}
\vspace{-5mm}
} %caption 
\end{figure}

{\bf Network scenario and local datasets.} 
Our experiments feature a typical medium-scale edge scenario~\cite{in-edge-ai}, 
with 20~learning nodes connected with, and coordinated by, an edge-based server. 
Five out of 20 nodes belong to each of the following
four categories so that the total amount of data remains constant:
\begin{itemize} 
\item {\em gold}, having 500~samples each,
representing all 10~classes (i.e., articles of clothing) present in the
Fashion-MNIST dataset; 
\item {\em silver}, with 200~samples each, still
belonging to all classes; 
\item {\em bronze}, with 500~samples each,
belonging to only two classes per node; 
\item {\em garbage}, with
200~samples each, belonging to two classes. 
\end{itemize} 
With the exception of ``gold'' ones, nodes suffer from either low quantity or
low diversity, hence, low quality,
of local data. Our experiments establish a correlation between
the category of each node and its contribution to the learning process,
thus allowing us to identify the best strategies to decide whether and how to
integrate each node in the learning process.

{\bf Model-weighting and node-dropping strategies.} 
As discussed in \Sec{strats}, the weights assigned to local models
during the averaging phase (step~3 of \Fig{sequence}) can account for
the quality and/or quantity of their local data. Specifically, we
consider the following options: 
\begin{itemize} 
\item {\em uniform}: all
local models are given equal weight; 
\item {\em no.\,samples}: weights
are proportional to the number of samples in each local dataset; 
\item
{\em no.\,classes}: weights are proportional to the number of classes in
each local dataset; 
\item {\em entropy}:
weights are proportional to the {\em entropy} of local data, an information-theoretic metric expressing, intuitively, how difficult it is to predict the class of a randomly-chosen local sample.
\end{itemize} 
The ``samples'' strategy
accounts for the quantity of local data, the ``classes'' one for its
quality, and the ``entropy'' one for both. ``Uniform'', where all
weights are equal, is added as a benchmark.

For node-dropping, we assume that five learning nodes are dropped after
iteration~1, and compare the state-of-the-art strategy of dropping the
slowest nodes~\cite{neglia,jointframework,client-selection} against the
alternative one of dropping the nodes with the lowest weight. The
rationale behind the latter strategy is that weights are linked to how
significant the contribution that nodes can give to the learning process is, thus,
dropping the lowest-weight nodes can reduce learning times without
impairing the learning quality.

\begin{figure} 
\centering
\includegraphics[width=.95\columnwidth]{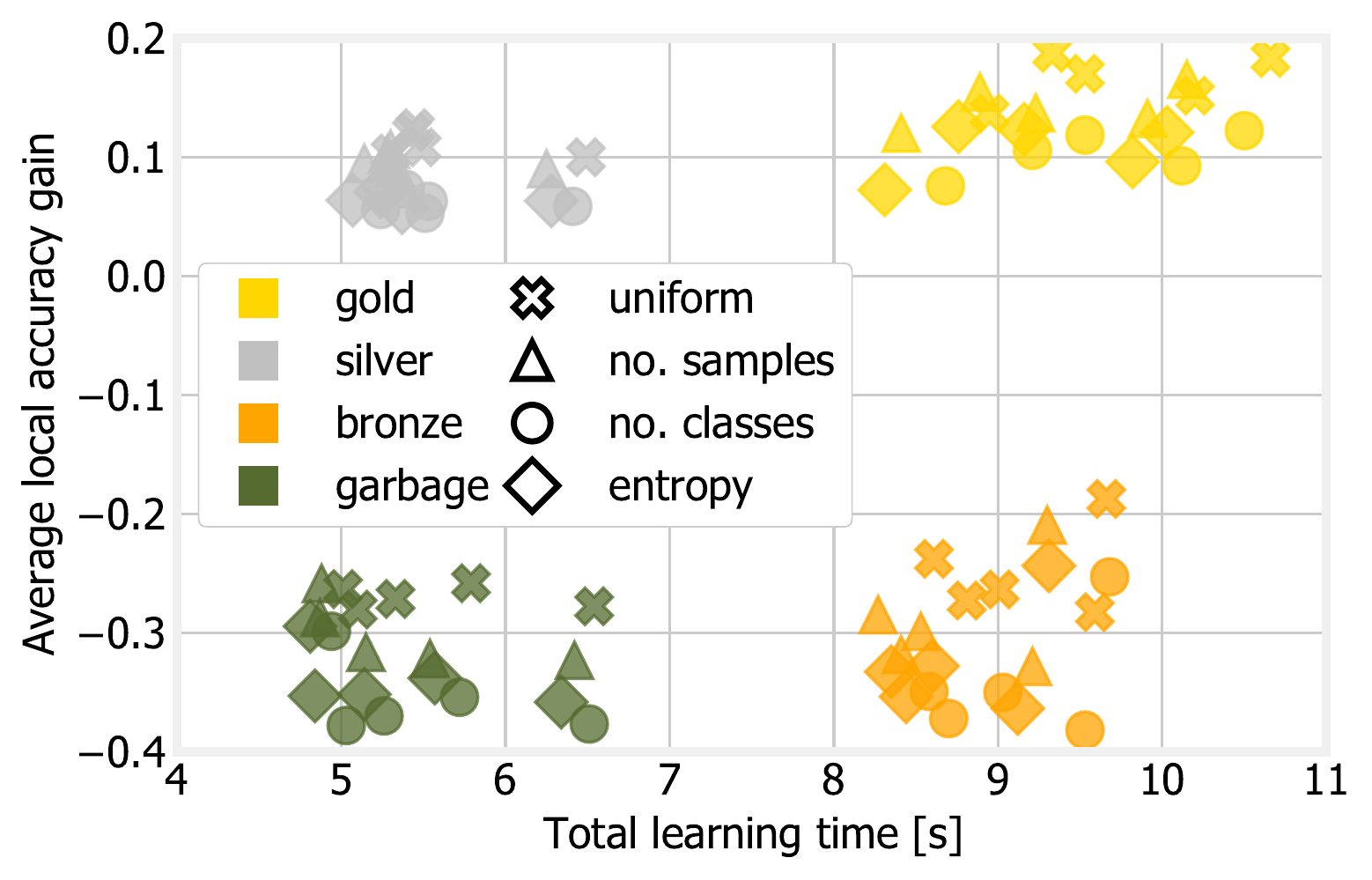}
\caption{Relationship between the average per-node learning time and
the local accuracy gain for different categories of nodes (identified by
the color of markers), under different model-weighting strategies (identified
by the shape of markers).\label{fig:timegain}} %caption 
\end{figure}

\subsection{Experiment results} \label{sec:sub-results}
The first aspect of interest is the progress of the overall learning,
i.e., the accuracy of the {\em global}, averaged model (step~3 of \Fig{sequence}),
portrayed in \Fig{learning}. Each marker therein represents the state of
the learning process after an iteration: its position along the x- and
y-axis represent (respectively) the elapsed time and the global
classification accuracy, while its color corresponds to the category of the
slowest node in that particular iteration. Different lines correspond to
different model-weighting strategies.

We can immediately observe that iteration times do not differ
substantially across model-weighting strategies, and that they are
usually determined by ``gold'' or ``bronze'' nodes -- which makes
intuitive sense, as those nodes have the largest local datasets. Even
more interestingly, we can observe a clear difference in the
classification accuracy obtained by different weighting strategies:
giving the same weight to all nodes, or only accounting for the size of
the local datasets, results in a lower accuracy than accounting for data
quality. Furthermore, there is little difference between the ``no.
classes'' and ``entropy'' strategies, suggesting that simply counting
the observed classes can be as effective as adopting more complex
metrics.

Next, \Fig{timegain} displays the relationship between the time taken by
local learning iterations and the
local accuracy gain, i.e., the improvement in classification accuracy obtained during local training (step~1 in \Fig{sequence}).
The latter metric can be seen as
a measure of how much individual nodes contribute to the global learning
process.
In the plot, the {\em color} of each marker represents the category of the corresponding node, while the {\em shape} of each marker represents the model-weighting strategy.
It is clear that, for all model-weighting strategies, local
learning times are strongly correlated with the quantity of local data,
while local accuracy gains are more strongly linked with the number of
classes, i.e., the data quality. These results also suggest that only
relying on local learning times for node-dropping decisions may result
in removing nodes with large, hence, potentially valuable, datasets.

\begin{figure} 
\centering
\includegraphics[width=.95\columnwidth]{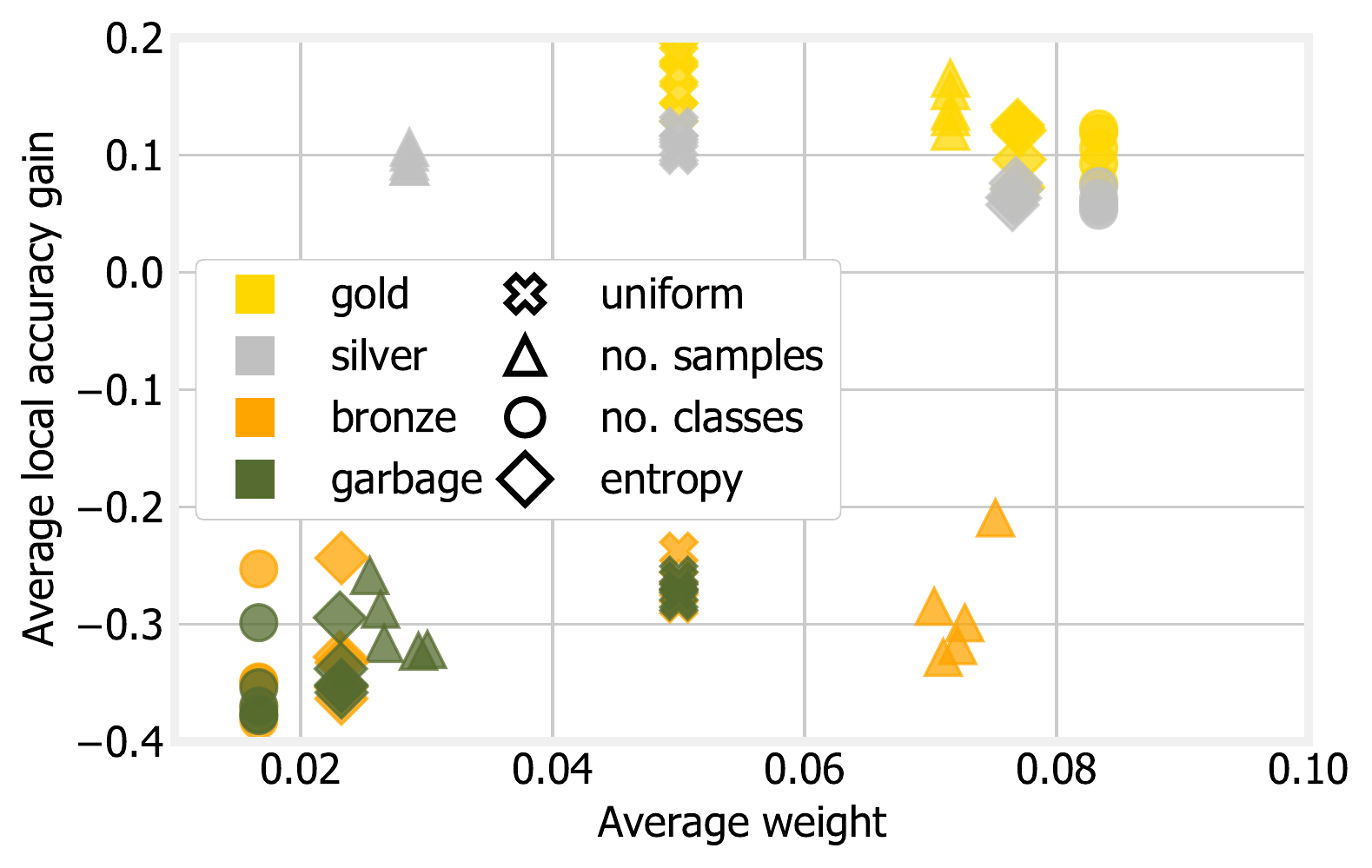}
\caption{Relationship between the weight assigned to local models and
the local accuracy gain for different categories of nodes (identified by
the color of markers), under different model-weighting strategies (identified
by the shape of markers). Correlation coefficients for the ``uniform'', ``no.
samples'', ``no. classes'' and ``entropy'' strategies are, respectively,
$0$, $0.07$, $0.98$ and $0.99$.\label{fig:weightgain}} %caption
\end{figure}

\Fig{weightgain} shows the relationship between the weight assigned to
each local model and the corresponding local accuracy gain.
Similar to \Fig{timegain}, the color and shape of markers represent, respectively, the node category and model-weighting strategy.
We quantify the relationship between weights and accuracy gains, by {\em correlation coefficients}, 
expressing to which extent changes in one  quantity are reflected by changes in the other: 
values close to~$1$ indicate strong correlation, values close to~$0$ little to no correlation.

In our case,
it is clear that
weights only considering the quantity of data (i.e., the ``no. samples''
strategy) may not be able to identify the nodes that can contribute the
most to the learning process, e.g., it gives similar weights to the
``silver'' and ``garbage'' nodes. On the other hand, ``no. classes'' and
``entropy'' weights are very well correlated with accuracy gains, which
again suggests how the quality of data has a high impact on
the learning effectiveness.
From both \Fig{timegain} and \Fig{weightgain}, it is also possible to see how better local accuracy gains do not necessarily coincide with better global classification accuracy. Indeed, good model-weighting decisions are necessary to consolidate local learning from different nodes into a consistent, high-quality global model.

\begin{figure} \centering
\includegraphics[width=.95\columnwidth]{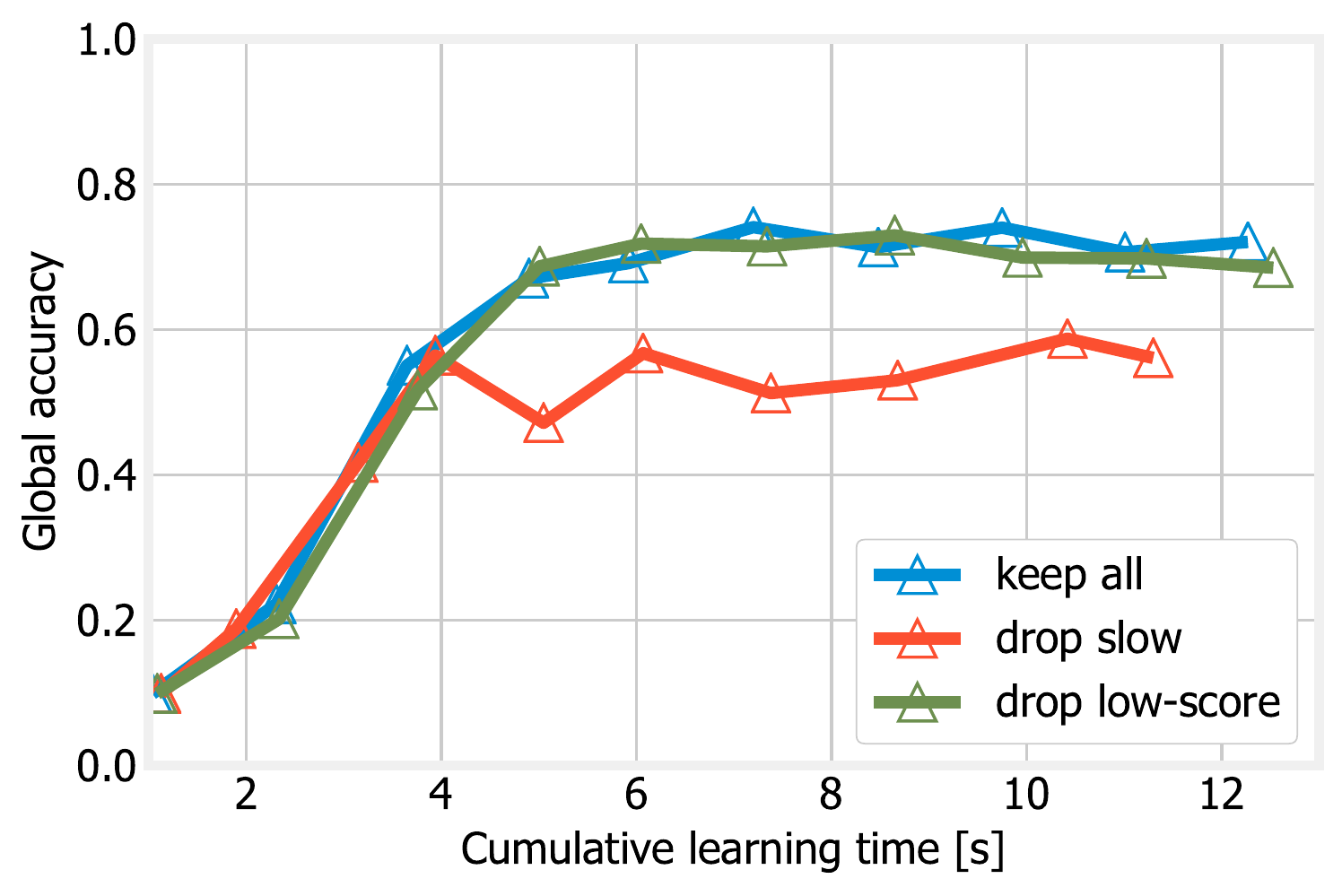}\\
\includegraphics[width=.95\columnwidth]{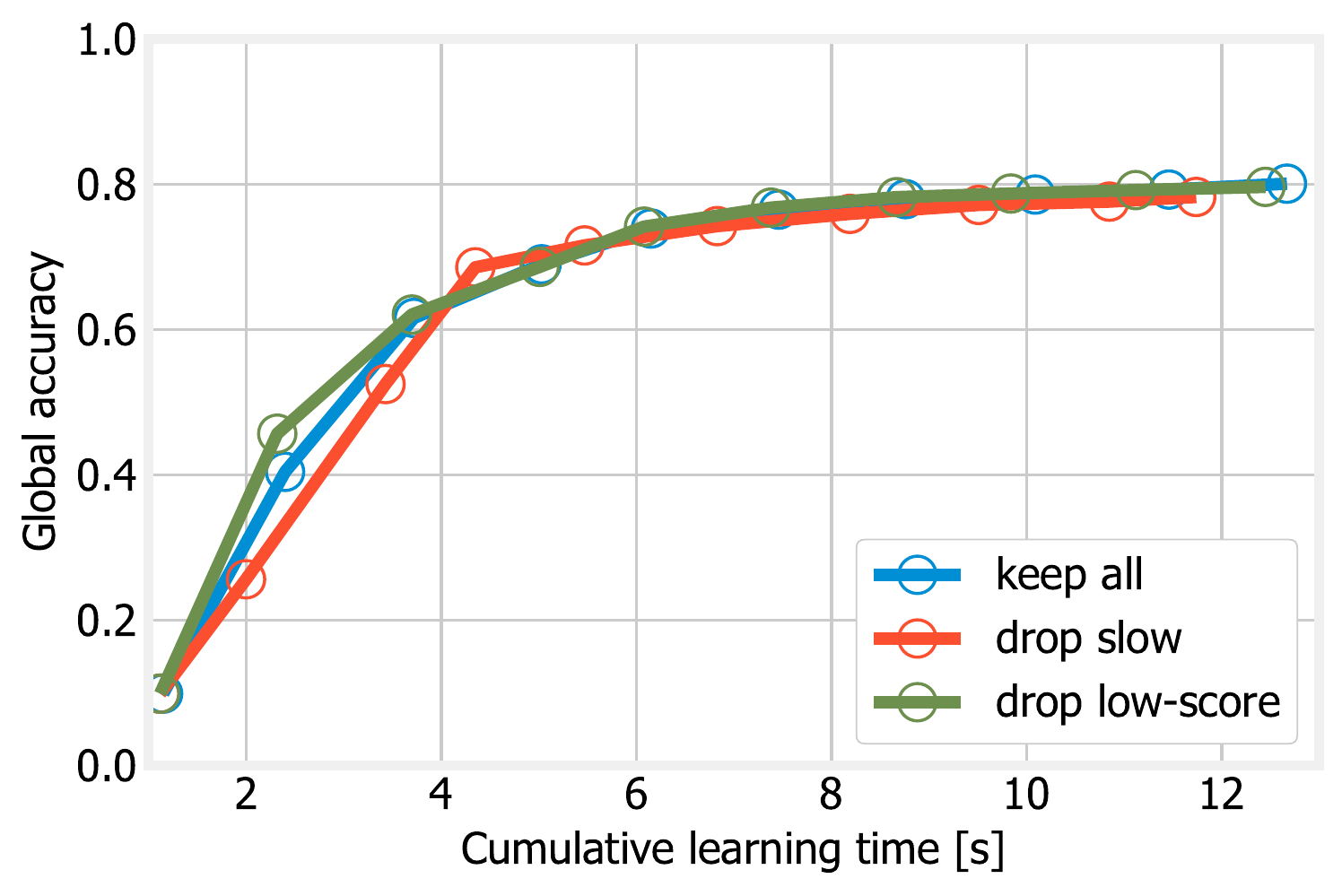}
\caption{Relationship between elapsed time and global accuracy for
different node-dropping schemes, under the ``no. samples'' (top) and
``no. classes'' (bottom) model-weighting strategies. 
\label{fig:drop}
\vspace{-5mm}
}%caption 
\end{figure}

Last, \Fig{drop} shows the effect of different node-dropping strategies
on the learning process, for the ``no. samples' and ``no. classes''
model-weighting strategies.
Specifically, we wait for the first five iterations, and then drop the five nodes with the lowest score, computed according to each node-dropping strategy.
We can observe that, when weights only
reflect the quantity of local data, dropping the slowest nodes
significantly hurts the learning accuracy. On the other hand, more
sophisticate model-weighting {\em or} node-dropping strategies yield
virtually the same accuracy as keeping all nodes. This also highlights
how  model-weighting and node-dropping decisions interact with one
another, and can represent different, complementary ways to achieve the
same goals.

{\bf Summary.}
In conclusion, our results show that the {\em quantity} of data drives the computation 
time of local nodes; however, it is the {\em quality} of data that determines 
its usefulness to the global learning process. 
It is thus of paramount importance that model-weighting decisions do not solely account 
for data quantity or computation time, as that may adversely impact performance.

\section{Take-away Messages and Challenges} \label{sec:takeaway}

Based on both the existing works discussed in \Sec{fl} and \Sec{strats}
and the experiments reported in \Sec{experiments}, we can highlight the
following high-level lessons learned, which also point at interesting
directions for future research.

\paragraph*{Model weights matter}

Assigning the right weights to local nodes during the averaging phase
can have a very significant impact on the learning process, as
highlighted in \Fig{learning} and \Fig{drop}.
Although few works in the literature have explored this option, \Fig{weightgain} shows how
\begin{comment}
Very few existing schemes
have deviated from the simple approach of making weights proportional to
the quantity of available local data; however, our experiments have
shown that more comprehensive metrics can yield much better performance.
Indeed, as highlighted in
\end{comment}
weights accounting for the
quality of data as well as its quantity are much more likely to
identify the nodes that can contribute the most to the learning process.
Importantly, the information needed to compute such weights is either
already available or easy to collect for the learning server, and does
not jeopardize the privacy properties of FL.
\begin{comment}
Defining and measuring data
quality, as well as integrating such figures with FL, are all novel and
potentially promising research challenges.
\end{comment}

\paragraph*{Data quality matters}

Our experiments strongly underline the importance of dataset quality.
An example is provided in \Fig{timegain} and
\Fig{weightgain}, showing how nodes with more diverse data (``gold'' and
``silver'') are able to offer much greater contributions to the overall
learning process.
Quantifying data quality is not a trivial task, however, our experiments show that even simple definitions based on counting the classes present in a given dataset yield very good results. Further research can explore additional
\begin{comment}
This implies that data quality shall be appropriately
estimated -- the number of represented classes is arguably the simplest
metric, but has proven very effective nonetheless -- and accounted for
when making model-weighting and node-dropping decisions. It is almost a
truism to say that both quality and quantity matter, and that both shall
be accounted for. However, our experiments suggest that quality may be
the most significant of the two aspects.

Further research could aim at
narrowing down the
\end{comment}
aspects of data quality, e.g., its freshness~\cite{alaa}.

%, having
%the most significant aspect on the learning performance.

\paragraph*{Check why nodes straggle before dropping them}

The global learning time of FL depends on the slowest node in each
iteration; therefore, it is often tempting to try and speed learning up
by dropping the slowest nodes~\cite{neglia}. Such a strategy is
appropriate when the slowest nodes are indeed stragglers, with limited
computational capabilities or poor
connectivity~\cite{jointframework,client-selection}; however, this may
result in unduly excluding nodes with valuable, rich datasets.
Privacy concerns often prevent the server from obtaining additional
information on individual learning nodes; however, already-available
data like the size of local datasets can provide significant help to 
tell genuine stragglers apart from nodes that simply have a lot of
data. If warranted, the latter can be directed to sample their own
datasets, in a similar spirit to~\cite{infocom20-noniid}, so as to
provide good contributions to the learning process with a smaller
latency. This also points at the exciting research direction of
extending the FL paradigm by allowing additional interaction between the
learning server and learning nodes, striking the right balance between
simplicity, privacy, and effectiveness.

\paragraph*{FL is robust}

This is not very surprising, since FL has been introduced for the very
purpose of exploiting local, potentially heterogeneous, data from
devices that cannot be centrally controlled. It is however interesting
to highlight how the robustness of FL extends beyond tackling
low-quality data, to tackling suboptimal configurations. An example is
provided in \Fig{drop}, where it is sufficient to make high-quality
model-weighting {\em or} node-dropping decisions to obtain very good
learning performance. This suggests that FL is indeed a viable choice in
those environments and scenarios where there is a significant likelihood
that configuration decisions be suboptimal. Robustness to incorrect
configuration is a relevant research area in distributed computing
scenarios, and -- so far -- a neglected one.

\section{Conclusion} \label{sec:conclusion}

In the context of federated learning, we have considered the problems of
model-weighting, i.e., assigning weights to local models during the
averaging phase, and node-dropping, i.e., selecting the nodes to exclude
from the learning process. 
After observing how those two decisions can
tackle most of the issues and hurdles of FL, we have reviewed existing
approaches thereto and found them to seldom depart from straightforward
solutions based on the quantity of local data learning nodes have and
their response time. Leveraging a set of real-world experiments, we have
observed how more comprehensive approaches, accounting for the quality
of local data and for the reasons behind longer node latency, can yield
substantially better learning performance.

\bibliographystyle{IEEEtran} 
\bibliography{refs}%

\begin{thebibliography}{10}
\providecommand{\url}[1]{#1}
\csname url@samestyle\endcsname
\providecommand{\newblock}{\relax}
\providecommand{\bibinfo}[2]{#2}
\providecommand{\BIBentrySTDinterwordspacing}{\spaceskip=0pt\relax}
\providecommand{\BIBentryALTinterwordstretchfactor}{4}
\providecommand{\BIBentryALTinterwordspacing}{\spaceskip=\fontdimen2\font plus
\BIBentryALTinterwordstretchfactor\fontdimen3\font minus
  \fontdimen4\font\relax}
\providecommand{\BIBforeignlanguage}[2]{{%
\expandafter\ifx\csname l@#1\endcsname\relax
\typeout{** WARNING: IEEEtran.bst: No hyphenation pattern has been}%
\typeout{** loaded for the language `#1'. Using the pattern for}%
\typeout{** the default language instead.}%
\else
\language=\csname l@#1\endcsname
\fi
#2}}
\providecommand{\BIBdecl}{\relax}
\BIBdecl

\bibitem{konen2015federatedOptimization}
J.~Konečný, B.~McMahan, and D.~Ramage, ``Federated optimization: Distributed
  optimization beyond the datacenter,'' \emph{arXiv preprint arXiv:1511.03575},
  2015.

\bibitem{jointframework}
M.~{Chen}, Z.~{Yang}, W.~{Saad}, C.~{Yin}, H.~V. {Poor}, and S.~{Cui}, ``A
  joint learning and communications framework for federated learning over
  wireless networks,'' \emph{IEEE Transactions on Wireless Communications},
  vol.~20, no.~1, pp. 269--283, 2020.

\bibitem{kang2020reliable}
J.~Kang, Z.~Xiong, D.~Niyato, Y.~Zou, Y.~Zhang, and M.~Guizani, ``Reliable
  federated learning for mobile networks,'' \emph{IEEE Wireless
  Communications}, vol.~27, no.~2, pp. 72--80, 2020.

\bibitem{infocom20-noniid}
H.~Wang, Z.~Kaplan, D.~Niu, and B.~Li, ``Optimizing federated learning on
  non-iid data with reinforcement learning,'' in \emph{IEEE INFOCOM}, 2020.

\bibitem{client-selection}
T.~{Nishio} and R.~{Yonetani}, ``{Client Selection for Federated Learning with
  Heterogeneous Resources in Mobile Edge},'' in \emph{IEEE ICC 2019}, 2019.

\bibitem{in-edge-ai}
X.~{Wang}, Y.~{Han}, C.~{Wang}, Q.~{Zhao}, X.~{Chen}, and M.~{Chen}, ``{In-Edge
  AI: Intelligentizing Mobile Edge Computing, Caching and Communication by
  Federated Learning},'' \emph{IEEE Network}, vol.~33, no.~5, pp. 156--165,
  2019.

\bibitem{neglia}
G.~{Neglia}, G.~{Calbi}, D.~{Towsley}, and G.~{Vardoyan}, ``The role of network
  topology for distributed machine learning,'' in \emph{IEEE INFOCOM}, 2019.

\bibitem{infocom20-fog}
Y.~{Tu}, Y.~{Ruan}, S.~{Wagle}, C.~G. {Brinton}, and C.~{Joe-Wong},
  ``{Network-Aware Optimization of Distributed Learning for Fog Computing},''
  in \emph{IEEE INFOCOM}, 2020.

\bibitem{robust-efficient}
F.~{Sattler}, S.~{Wiedemann}, K.-R. {M\"{u}ller}, and W.~{Samek}, ``{Robust and
  Communication-Efficient Federated Learning From Non-i.i.d. Data},''
  \emph{IEEE Transactions on Neural Networks and Learning Systems}, vol.~31,
  no.~9, pp. 3400--3413, 2020.

\bibitem{wang2019adaptive}
S.~Wang, T.~Tuor, T.~Salonidis, K.~K. Leung, C.~Makaya, T.~He, and K.~Chan,
  ``Adaptive federated learning in resource constrained edge computing
  systems,'' \emph{IEEE Journal on Selected Areas in Communications}, vol.~37,
  no.~6, pp. 1205--1221, 2019.

\bibitem{lu2019differentially}
Y.~Lu, X.~Huang, Y.~Dai, S.~Maharjan, and Y.~Zhang, ``Differentially private
  asynchronous federated learning for mobile edge computing in urban
  informatics,'' \emph{IEEE Transactions on Industrial Informatics}, vol.~16,
  no.~3, pp. 2134--2143, 2019.

\bibitem{shin2020xor}
M.~Shin, C.~Hwang, J.~Kim, J.~Park, M.~Bennis, and S.-L. Kim, ``{XOR Mixup:
  Privacy-Preserving Data Augmentation for One-Shot Federated Learning},''
  \emph{arXiv preprint arXiv:2006.05148}, 2020.

\bibitem{plos-entropy}
H.~Huang, J.~Huang, Y.~Feng, J.~Zhang, Z.~Liu, Q.~Wang, and L.~Chen, ``On the
  improvement of reinforcement active learning with the involvement of cross
  entropy to address one-shot learning problem,'' \emph{PloS one}, vol.~14,
  no.~6, p. e0217408, 2019.

\bibitem{alaa}
A.~A. Abdellatif, C.~F. Chiasserini, and F.~Malandrino, ``Active learning-based
  classification in automated connected vehicles,'' in \emph{IEEE INFOCOM
  PERSIST-IoT Workshop}, 2020.

\end{thebibliography}
\vskip -1cm plus -1fil
\begin{IEEEbiographynophoto}%[{\includegraphics[width=1in,height=1.25in,clip,keepaspectratio]{photos/francesco.jpg}}]%
{Francesco Malandrino}
(M'09, SM'19) earned his Ph.D. degree from Politecnico di Torino in 2012
and is now a researcher at the National Research Council of Italy
(CNR-IEIIT). His research interests include the architecture and
management of wireless, cellular, and vehicular networks.
\end{IEEEbiographynophoto}
\vskip -1cm plus -1fil
\begin{IEEEbiographynophoto}%[{\includegraphics[width=1in,height=1.25in,clip,keepaspectratio]{photos/carla.jpg}}]%
{Carla~Fabiana Chiasserini}
(M'98, SM'09, F'18) received her Ph.D. from Politecnico di Torino in
2000. She is currently a Full Professor  with  the  Department  of
Electronic  Engineering  and  Telecommunications at Politecnico di
Torino, as well as  the Vice Rector for Alumni and Career Orientation. 
Her research interests include architectures, protocols, and performance
analysis of wireless networks. 
\end{IEEEbiographynophoto}

\end{document}